\documentclass[10pt,conference]{IEEEtran}
\usepackage{lscape}
\usepackage{amsfonts,syntonly}
\usepackage{verbatim,cite,amsmath}
\usepackage{array}
\usepackage{color}
\usepackage{graphicx}
\usepackage{amsmath}
\usepackage{flushend}
\usepackage{stfloats}

\usepackage[latin1]{inputenc}

\newtheorem{theorem}{Theorem}
\newtheorem{remark}{Remark}
\newtheorem{example}{Example}

\DeclareMathOperator{\E}{E}
\DeclareMathOperator{\Prob}{Prob}

\begin{document}

\title{DMT of Weighted Parallel Channels: Application to Broadcast Channels}

%\author{\authorblockN{Lina Mroueh}
%\authorblockA{Télécom ParisTech\\
%75013 Paris, France\\
%mroueh@enst.fr}
%\and
%\authorblockN{Stéphanie Rouquette-Léveil}
%\authorblockA{Motorola Labs\\
%91193, Gif sur Yvette, France\\
%stephanie.rouquette@motorola.com}
%\and
%\authorblockN{Ghaya Rekaya-Ben Othman}
%\authorblockA{Télécom ParisTech\\
%75013 Paris, France\\
%rekaya@enst.fr}
%\and
%\authorblockN{Jean-Claude Belfiore}
%\authorblockA{Télécom ParisTech\\
%75013 Paris, France\\
%belfiore@enst.fr}
%}

% avoiding spaces at the end of the author lines is not a problem with
% conference papers because we don't use \thanks or \IEEEmembership
\author{\authorblockN{Lina Mroueh\authorrefmark{2}\authorrefmark{1},
Stéphanie Rouquette-Léveil\authorrefmark{2},
Ghaya Rekaya-Ben Othman\authorrefmark{1} and
Jean-Claude Belfiore\authorrefmark{1}}
\authorblockA{\authorrefmark{1}
Télécom ParisTech, 75013 Paris, France \\ Email: \{mroueh, rekaya, belfiore\}@enst.fr}
\authorblockA{\authorrefmark{2}
Motorola Labs, Gif sur Yvette, France \\
Email: stephanie.rouquette@motorola.com}
}

\maketitle

\begin{abstract}
In a broadcast channel with random packet arrival and transmission queues, the stability of the system is achieved by maximizing a weighted sum rate capacity with suitable weights that depend on the queue size. The weighted sum rate capacity using Dirty Paper Coding (DPC) and Zero Forcing (ZF) is asymptotically equivalent to the weighted sum capacity over parallel single-channels. 
In this paper, we study the Diversity Multiplexing Tradeoff (DMT) of the fading broadcast channel under a fixed weighted sum rate capacity constraint. The DMT of both identical and different parallel weighted MISO channels is first derived. Finally, we deduce the DMT of a broadcast channel using DPC and ZF precoders. % However, the maximal diversity and multiplexing gain are independent from the weight distribution. We find also the weights that maximize the DMT.
\end{abstract}

\section{Introduction and Motivations}
The multiple antenna broadcast channel (BC) has recently gained attention, due to the fact that this channel can provide MIMO spatial multiplexing benefits without requiring multiple antennas at the receiver. It is well-known that the Dirty Paper Coding (DPC) achieves the maximum sum capacity. However, the implementation of DPC brings high complexity to both the transmitter and the receiver. As the capacity-achieving dirty paper coding approach is difficult to implement, many more practical downlink transmission techniques have been proposed. Downlink linear beamforming, although suboptimal, has been shown to achieve a large portion of DPC capacity while being simpler to operate than DPC~\cite{Peel-ZF-MMSE}.

\begin{figure}[ht]
\centering
 \includegraphics[width = 0.9\linewidth]{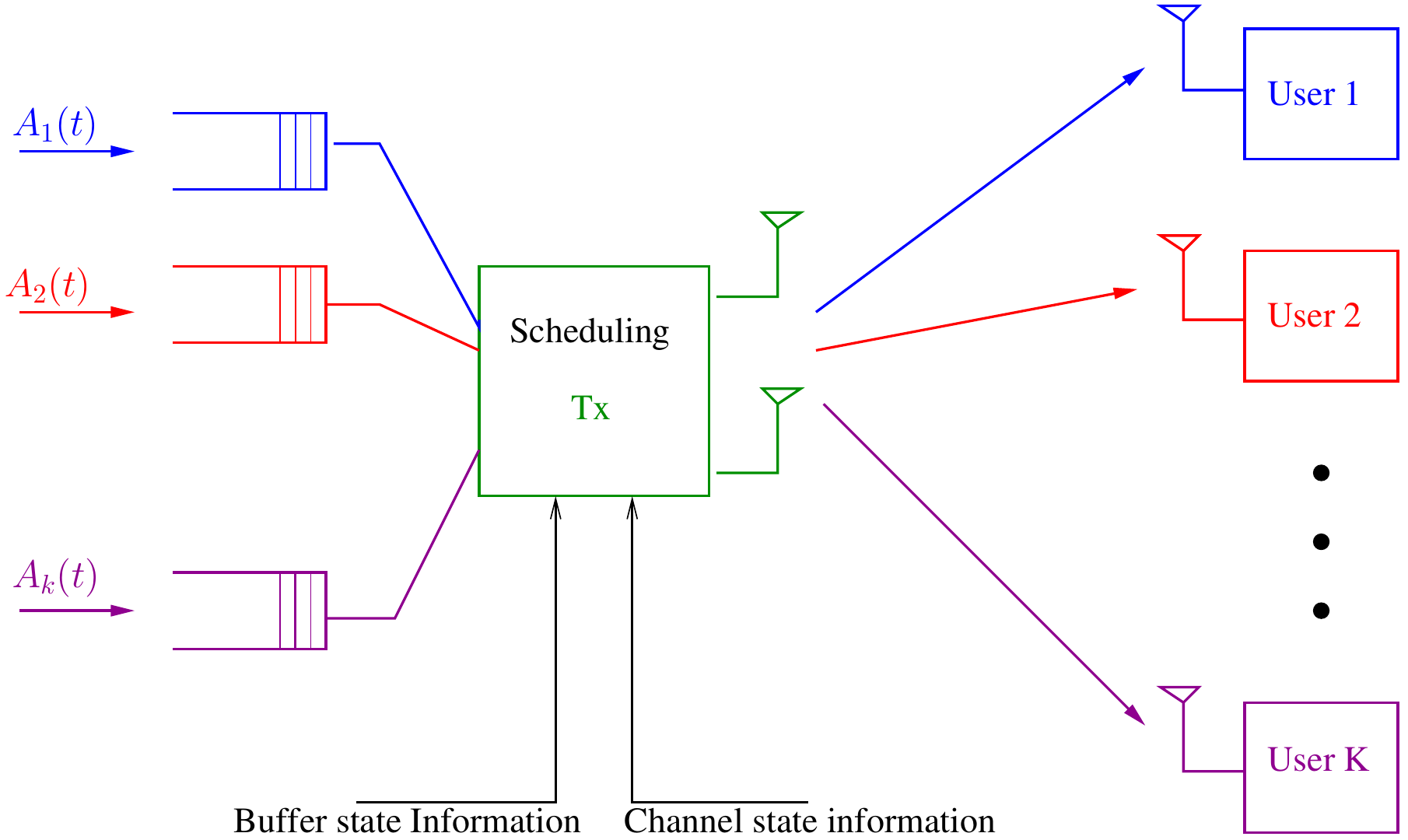}
 \caption{Broadcast channel with random packet arrivals and transmission queues}
 \label{fig:bc-buffer}
\end{figure}

In modern packet-based wireless data networks, traffic arrives at random time instants at the transmitter in the form of variable-sized packets. In queuing system, the stability region is the set of all bit arrival rates for which no queue size blows up. It has been shown in \cite{Yeh-MAC-PHY} that under random packet arrival and transmission queues, the system stability is achieved by maximizing a weighted sum rate capacity with suitable weights $\mu_i$ that depend on the queue size. In this case, joint power control and rate allocation should be investigated based on both buffer state information and channel state information as illustrated in fig.~\ref{fig:bc-buffer}. The stability of the system is therefore guaranteed for a given weighted sum rate with suitable rate and power allocation policy. Whenever transmission with a given rate $R$ that exceed this weighted sum rate capacity occurs, the system is unstable. The probability of being unstable is what we call in the following \emph{outage probability}. On this outage performance, there is a tradeoff: The higher the required weighted sum rate is, the lower the reliability of the system is and vice-versa.   
The outage probability is then defined as 
\begin{equation}
P_{\text{out}}(R) = \Prob\big\{K \displaystyle \sum_{i=1}^{K} \mu_i R_{i} \leq R \big\}
\end{equation}
where $R$ is the fixed transmission rate, $K$ the number of users and $\sum_i \mu_i = 1$.

At asymptotically high SNRs, it was shown in \cite{Jindal-high-DPC,Jindal-high-DPC-1} that the weighted sum rate maximization problem for DPC and ZF strategies can be decoupled into maximization over independent parallel channels. The asymptotic power allocation that maximizes the weighted sum capacity consists therefore on allocating power proportionally to user weights \cite{Jindal-high-DPC,Jindal-high-DPC-1}. 
Our objective in this paper is to analyse the outage Diversity Multiplexing Tradeoff (DMT) under fixed weighted sum rate capacity constraint. 

%v
%The system performance can be improved by assigning variable rates $\mu_i R_i$ to the $K$ different users, with $\sum_i \mu_i = 1$. In this case, individual rates are assigned with respect to the buffer and to the channel state information as illustrated in fig.~\ref{fig:bc-buffer}.

%In high SNR regime, a convenient characterization of the tradeoff between rate and reliability is offered by the DMT introduced by Zheng and Tse in~\cite{DMT-Zheng-MIMO}. 
%Given a point-to-point MIMO system, the gains in terms of diversity gain $d$ 
%\begin{equation*}
%-d = \displaystyle \lim_{\SNR \rightarrow \infty} \frac{P_{\text{out}}(\SNR)}{\log\SNR}
%\end{equation*}
%and spatial multiplexing gain $r$
%\begin{equation*}
%r = \displaystyle \lim_{\SNR \rightarrow \infty} \frac{R(\SNR)}{\log\SNR}
%\end{equation*}
%can be simultaneously obtained. But, there is a fundamental tradeoff between these two gains provided by any coding scheme.

%In MIMO broadcast channel, when full CSIT is assumed at the transmitter, without any assumption on the transmission rates, the outage analysis is not meaningful. In this paper, we are interested in a \emph{fixed weighted sum rate} $R$ instead of individual rates. 

As the weighted sum rate capacity of DPC and linear precoding is asymptotically equivalent to the weighted sum capacity over parallel single-channels \cite{Jindal-high-DPC,Jindal-high-DPC-1}, we first study the DMT of weighted parallel channels. Then, we apply this DMT to the broadcast case. The remainder of this paper is organized as follows. In section \ref{sec:background}, we define the system model and we provide background material on weighted sum capacity analysis at high SNRs. We derive in section \ref{sec:Weight-MISO} the DMT of parallel weighted channels when identical and different parallel channels are considered. In section \ref{sec:broadcast}, we deduce from section \ref{sec:Weight-MISO} the DMT of the broadcast case using ZF and DPC. Finally, section \ref{sec:conclusion} concludes this paper. 

\section{System Model and Preliminaries}
\label{sec:background}
\subsection{System Model}
We consider a K receiver multiple-antenna broadcast channel in which the transmitter has $M$ antennas and each receiver has a single antenna, with $M \geq K$.
The received signal $y_k$ for user $k$ is given by
\begin{equation*}
y_k = \mathbf{h}_k \mathbf{x} + n_k \quad k = 1,\ldots,K
\end{equation*} 
where $\mathbf{h}_1,\mathbf{h}_2,\ldots, \mathbf{h}_K$ are the channel vectors (with $\mathbf h_i \in \mathbb C^{1 \times M}$) of user $1$ to $K$, with i.i.d unit variance Gaussian entries. The vector $\mathbf x \in \mathbb C^{1 \times M}$ is the transmitted signal, and $n_1, \ldots, n_k$ are independent complex Gaussian noise terms with unit variance. The input must satisfy a transmit power constraint of $\rho$, \textit{i.e.}, $\E[\|x\|^2] \leq \rho$. We assume that the transmitter has perfect knowledge of the whole channel matrix and each receiver has perfect knowledge of its own equivalent channel matrix. 

\subsection{Weighted sum rate analysis at high SNR} 
\label{sec:weighted-sum}
In \cite{Jindal-high-DPC}, it has been shown that under total power constraint $\rho$, the weighted sum rate capacity for Zero Forcing (ZF) and DPC is asymptotically equivalent to the weighted sum capacity over parallel single-channels. The asymptotical power allocation that maximizes the weighted sum capacity consists on allocating the power proportionally to user weights.
\\*
As shown in \cite{Jindal-high-DPC, Jindal-high-DPC-1}, the weighted sum rates of Zero Forcing (ZF) and DPC at high SNR can be expressed as 
\begin{equation}
C_{\text{ZF}}(\boldsymbol{\mu},\mathbf{H},\rho) \cong \displaystyle \sum_{i=1}^{K} \mu_i \log\big(1 + \mu_i \rho \|g_i\|^2\big) 
\label{eq:cap_ZF}
\end{equation}
\begin{equation}
C_{\text{DPC}}(\boldsymbol{\mu},\mathbf{H},\rho) \cong \displaystyle \sum_{i=1}^{K} \mu_i \log\big(1 + \mu_i \rho \|f_i\|^2\big) 
\label{eq:cap_DPC}
\end{equation}
$\cong$ refers to equivalence in the limit ($\rho \rightarrow \infty$).
\\*
$g_i$ is the projection onto the null space of $\{\mathbf h_1, \ldots,\mathbf h_{i-1}, \mathbf h_{i+1}, \ldots, \mathbf h_K \}$. $f_i$ is the projection onto the null space of $\{\mathbf h_1, \ldots,\mathbf h_{i-1}\}$. 
\\*
In Rayleigh fading, the distributions of $\|g_i\|^2$ and $\|f_i\|^2$ are $\chi_{2(M-K+1)}^2$ and $\chi_{2(M-i+1)}^2$ respectively. 
\\*
As the weighted sum rate capacity of ZF and DPC is asymptotically equivalent to the weighted sum capacity over parallel single-channels, we study in this paper first the DMT of weighted parallel channels. Then, we deduce the DMT of the broadcast channel using these techniques.
%
%\subsection{Outage definition}

%In this section, we recall briefly the DMT of $n_t \times n_r$ MIMO channel as defined in \cite{DMT-Zheng-MIMO}. 
%\begin{theorem}
%For a point-to-point MIMO system, the DMT of a MIMO channel \cite{DMT-Zheng-MIMO} is a linear-piecewise function connecting the points $(k,d(k))$ where $$d(k)= (n_r - k)(n_t -k)$$ and $ k = 0 , \ldots, q = \min\{n_t,n_r\}$, 
%\end{theorem} 
%
%\textit{Proof:} From \cite{DMT-Zheng-MIMO}, we have
%\begin{equation*}
%d_{n_t,n_r}(r) = \inf_{\mathcal O(\boldsymbol{\alpha})} \displaystyle \sum_{i=1}^{q}(2i-1 + |n_t-n_r|)\alpha_i
%\end{equation*}
%with $\boldsymbol{\alpha}$ being the ordered exponential orders of the $q$ eigen values of the channel matrix. $\mathcal{O}$ define the typical outage event, and is given by 
%\begin{equation*}
%\mathcal{O}(\boldsymbol{\alpha}) = \Big\{\boldsymbol{\alpha} \in \mathbb R^q_+ : \displaystyle \sum_{i=1}^{q}(1 - \alpha_i)^+ < r \Big\}
%\end{equation*}

\section{DMT of weighted parallel MISO or SIMO channels}
\label{sec:Weight-MISO}
In this section, we study the DMT of MISO channels when weights are affected to different channels. We study separately the cases of identical and different parallel channels. Note that the DMT when SIMO parallel channels are considered is the same as the MISO case.  

\subsection{DMT of K parallel identical MISO channels} 
\begin{theorem}
\label{theo:DMT-sym}
The DMT of K parallel $n_t\times 1$ MISO channels under a total power constraint $\rho$ when different weights ($\mu_1 \geq \ldots \geq \mu_K > 0$, with $\sum_i \mu_i = 1$) are affected to the channels is a piecewise-linear function connecting the points $(r(i),d(i))$, $i = 0, \ldots, K$ where 
\begin{equation}
\begin{array}{lll}
r(i) = K\Big(1 - \displaystyle \sum_{j=1}^{K - i} \mu_j\Big) & 0 \leq i \leq  K - 1; & r(K) = K \\
d(i) = n_t(K-i) & 0 \leq i \leq  K  & \\
\end{array}
\label{eq:DMT_ident}
\end{equation} 
\label{theo:DMT_ident}
\end{theorem}
Before going to the proof, some remarks can be made about this DMT.

\begin{remark}
From (\ref{eq:DMT-1}), we note that
\begin{enumerate}
\item
The maximal diversity $d^*_{\text{out}}(0) = Kn_t$ is independent from the weight distribution.
\item
When $\mu_1 = \ldots = \mu_K = \frac{1}{K}$, the DMT is $d^*(r,\text{uniform}) = Kn_t(1-r/K)$, which corresponds to the DMT calculated in \cite{Sheng-Par-Chan} for the uniform case. $d^*(r,\text{uniform})$ is the upperbound of $d(r,\mu)$. The upperbound corresponds to the case when all the channels are in outage with a target rate $\frac{R}{K}$ for every channel. 
\end{enumerate}
\end{remark}

\begin{figure*}[!b]
\hrulefill
\begin{equation}
d^*_{\text{out}}(r,\mu) = 
\begin{cases}
\begin{array}{ll}
n_t\Big[ K-i-1 + \frac{1}{\mu_{K-i}}\big(1 - \frac{r}{K} - \displaystyle \sum_{j = 1}^{K - i -1} \mu_j\big) \Big] & r \in [r_i\;;\; r_{i+1}]\;,\;0 \leq i \leq K-2 \\
n_t \Big[\frac{1}{\mu_{1}}\big(1 - \frac{r}{K}\big) \Big] & r \in [r_{K-1}\;;\; r_{K}] \\
\end{array}
\label{eq:DMT-1}
\end{cases}
\end{equation}
\end{figure*}

\begin{proof}
The asymptotical power allocation that maximizes the weighted sum capacity consists in allocating the power proportionally to user weights \cite{Jindal-high-DPC}. 
The weighted sum rate capacity expression of the $K$ parallel channels is therefore 
\begin{equation*}
C_{K,n,1}\big(\boldsymbol{\mu},\mathbf{H}\big) \cong K \displaystyle \sum_{k=1}^{K} \mu_k \displaystyle \log\left(1 + \mu_k \rho\;h_{k}h_{k}^{H} \right) \\
\end{equation*}
$h_k$ is the $1 \times n_t$ MISO channel vector. $\lambda_{k} = h_{k}h_{k}^{H}$ is a Chi-squared variable with $n_t$ degrees of freedom, which is equivalent to $1\times1$ complex Wishart matrix with $n_t$ degrees of freedom.  If $\lambda_{k} = \rho^{-\alpha_{k}}$, then at high SNR, $1+\mu_k\rho\lambda_{k}~\sim~\mu_k\rho^{(1 - \alpha_{k})^+}$.
\\*
The outage probability can be written as
$$ P_{out}(r\log \rho) = \Prob\big\{C_{K,n,1}\big(\boldsymbol{\mu},\mathbf{H}\big) \leq r\log\rho \big\} $$
Let $\mathcal{O}$ describes the outage event, then 
\begin{equation*}
\mathcal{O} = \Big\{ \boldsymbol{\alpha} \in \mathbb R^K_+: \displaystyle \sum_{k=1}^{K} \mu_k \left[ \log\left(\mu_k \rho^{(1 - \alpha_{k})^+} \right) - \frac{r}{K} \log \rho \right] \leq 0 \Big\} \\
\end{equation*}
which is equivalent to 
$$
\mathcal{O} = \Big\{
\underbrace{\displaystyle \sum_{k=1}^{K} \mu_k \left[(1-\alpha_{k})^+ - \frac{r}{K} \right]}_{a} \log \rho + \underbrace{\displaystyle \sum_{k=1}^{K} \mu_k \log \mu_k}_{b}  \leq  0 \\
\Big\}
$$
At high SNR, $a \log \rho + b \leq 0 \Leftrightarrow a \leq 0$. Thus, the outage typical event could be written such as 
\begin{equation}
\mathcal{O} = \Big\{\alpha \in \mathbb R^K_+: \displaystyle \sum_{k=1}^{K} \mu_k \left[ (1-\alpha_{k})^+ - \frac{r}{K} \right] \leq 0\Big\}
\label{eq:outage}
\end{equation}
%with 
%\begin{equation}
%\mu_1 \geq \ldots \geq \mu_K > 0
%\label{eq:mu-const}
%\end{equation}
%\\*
The outage probability is 
$$
P_{\text{out}}(r\log\rho)   = \int_{\mathcal{O}} p(\alpha_1,\ldots,\alpha_K)\;d\alpha_1 \ldots d\alpha_K 
$$
Following \cite{DMT-Zheng-MIMO}, and using the fact that the channels are independent and identically distributed, the joint pdf is  
$$
p(\alpha_1,\ldots,\alpha_K) = p(\alpha_1)\ldots p(\alpha_K) \doteq \prod_{k=1}^{K} \rho^{-n_t\alpha_k}
$$
Finally, using Laplace's method as shown in \cite{DMT-Zheng-MIMO}
$$
P_{\text{out}}(r\log\rho) \doteq \int_{\mathcal{O}} \displaystyle \prod_{k=1}^{K} \rho^{-n_t\alpha_{k}} d\alpha_1 \ldots d\alpha_K = \rho^{-d_{\text{out}}(r)}
$$
where 
\begin{equation}
d_{\text{out}}(r,\boldsymbol{\mu}) = \inf_{(\alpha_1,\ldots,\alpha_K) \in \mathcal{O}}  n_t \displaystyle \sum_{k=1}^{K} \alpha_{k}
\label{eq:diversity}
\end{equation}

Note that all the $\alpha_i$ coefficients have the same contribution in the objective function. In addition, $\mu_i$ are ordered. Then, we can assume with out loss of generality of the optimal solution that $1 \geq \alpha_1 \geq \ldots \geq \alpha_K$.
The linear optimisation problem is therefore equivalent to the following problem 
\begin{equation*}
\begin{cases}
\text{Minimize: } \;\alpha_1 + \ldots + \alpha_K \\
\text{Such that: }\;\;0 \leq \alpha_{i} \leq 1 \quad \forall i\\ \
\mu_1 \alpha_1 + \ldots + \mu_K\alpha_K \geq 1 - \frac{r}{K} \\ 
\end{cases}
\label{eq:LP-1}
\end{equation*}
If $\underline{r = 0}$, the optimal solution is 
$$
\alpha_1^* = \ldots = \alpha_K^* = 1
$$
If $\underline{r \neq 0}$, then the optimal solution is 
\begin{equation}
\begin{array}{l}
\alpha_1^* = \min \Big[ \frac{1}{\mu_1}\Big(1 - \frac{r}{K}\Big)^+ , 1 \Big] \\
\alpha_{i}^* = \min \Big[ \frac{1}{\mu_i}\Big(1 - \frac{r}{K} - \displaystyle \sum_{j=1}^{i-1}\mu_j \Big)^+, 1 \Big] \quad \forall i \geq 2\\
\end{array}
\label{eq:LP-sol1}
\end{equation}
and the DMT is given by
\begin{equation*}
\begin{array}{l}
d^*_{\text{out}}(0) = Kn_t \\
d^*_{\text{out}}(r,\mu) = n_t\displaystyle \sum_{i=1}^{K} \alpha_i^* \\
\end{array}
\label{eq:DMT-2}
\end{equation*}
which is equivalent to the expression of the DMT given in (\ref{eq:DMT-1}). By replacing $r_i$ by its value in (\ref{eq:DMT-1}), we get the final DMT expression in (\ref{eq:DMT_ident}).
 
\end{proof}

\begin{example} We consider 2 weighted parallel $2\times1$ MISO channels. The DMT with respect to different weight distributions is illustrated in fig. \ref{fig:DMT_sym}.
\begin{figure}[ht]
\centering
 \includegraphics[width = 0.95\linewidth]{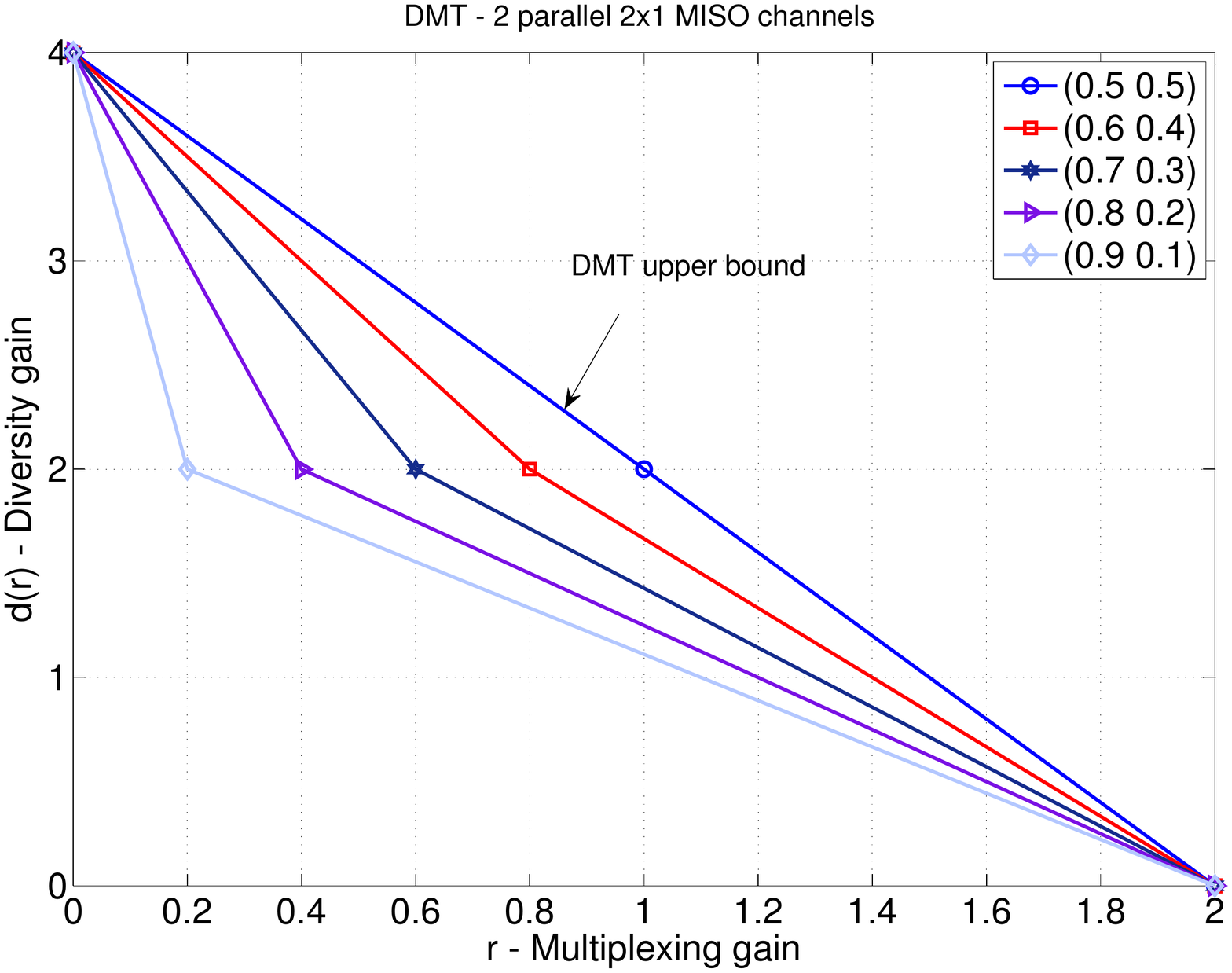}
\caption{DMT of 2 parallel identical $2 \times 1$ MISO channels with different weight distribution}
\label{fig:DMT_sym}
\end{figure}
We remark that, even if the extremal points ($d=0$ and $r=0$) do not depend on the weight distribution, the more unbalanced the weights are, the worse the DMT is. 
\end{example}
% As we can see, the DMT depends on the value of $\mu$. Plus les $mu_i$ sont equilibrés, plus la DMT est meilleure
% \textcolor{blue}{Lina has no idea about what should be in comments, Jean-Claude: commentaire ... }

\begin{figure*}[!b]
\hrulefill
\begin{equation}
d^*_{\text{out}}(r,\mu) = 
\begin{cases}
\begin{array}{ll}
\displaystyle \sum_{j=1}^{K-i-1} \hat{n_j} + \frac{\hat{n_{K-i}}}{\hat{\mu_{K-i}}}\big(1 - \frac{r}{K} - \displaystyle \sum_{j = 1}^{K - i -1} \hat{\mu_j}\big) \Big] & r \in [r_i\;;\; r_{i+1}]\;,\;0 \leq i \leq K-2 \\
\frac{\hat{n_1}}{\hat{\mu_{1}}}\big(1 - \frac{r}{K}\big)  & r \in [r_{K-1}\;;\; r_{K}] \\
\end{array}
\label{eq:DMT-asym}
\end{cases}
\end{equation}
\end{figure*}

\subsection{DMT of K parallel different MISO channels}
\begin{theorem}
We consider K parallel $n_{1}~\times~1,\ldots, n_{K}~\times~1$ MISO channels under a total power constraint $\rho$, where weights $\mu_1,\ldots,\mu_K$ are affected to different channels, with $\sum \mu_i = 1$ and $\mu_i > 0$, $\forall i$ .
Let us define $\bar{\mu_i}$ by 
$$
\bar\mu_i = \frac{\mu_i}{n_i}
$$ 
Let $\bar\mu_{i_1} \geq \ldots \geq \bar\mu_{i_K}$ be the ordered combination of $\bar\mu_i$, and $\hat{\mathbf {u}} = T(u)$ be the component combination of a vector $u$, such that $(\hat{u}_1,  \ldots, \hat{u}_K) =  (u_{i_1}, \ldots, u_{i_K})$.
\\*
Then, the diversity-multiplexing tradeoff is a piecewise linear function connecting the point $(r(i),d(i))$
where 
\begin{equation}
\begin{array}{lll}
r(i) = K\Big(1 - \displaystyle \sum_{j=1}^{K - i} \hat{\mu_j}\Big) & 0 \leq i \leq  K - 1; & r(K) = K \\
d(i) =  \displaystyle \sum_{j=1}^{K - i}\hat{n_j} & 0 \leq i \leq  K-1; &  d(K) = 0 \\
\end{array}
\label{eq:DMT_diff}
\end{equation} 
with $\hat{\mathbf{n}} = T(\mathbf{n})$ and $\hat{\boldsymbol{\mu}} = T(\boldsymbol{\mu})$.
\label{theo:DMT_dif}
\end{theorem}

\begin{remark} From the expression of the DMT, we note that
\begin{enumerate}
\item
The maximal diversity is independent of the weight distribution: $d(0) = \sum_i n_i$.
\item
The DMT is upperbounded by $d(r,\boldsymbol{\mu^*})$ where 
\begin{equation*}
\frac{\mu_1^*}{n_1} = \ldots = \frac{\mu_K^*}{n_K} = \frac{1}{d(0)}
\label{eq:mu}
\end{equation*}
In this case, the upperbound is given by
$$
d\big(r,\mu^*\big) = d(0)(1 - \frac{r}{K})
$$ 
This upperbound corresponds also to the case when all the channels are in outage with a target rate of $\frac{R}{K}$ for every channel. 
\end{enumerate}
\end{remark}

\begin{proof}
In this case, the $\alpha_i$ are not identically distributed. The joint distribution is 
therefore
$$
p(\alpha_1,\ldots,\alpha_K) = p(\alpha_1)\ldots p(\alpha_K) \doteq \prod_{k=1}^{K} \rho^{-n_k\alpha_k}
$$
The outage probability is 
\begin{eqnarray*}
P_{\text{out}}(r\log\rho) &=& \int_{\mathcal{O}} p(\alpha_1,\ldots,\alpha_K) d \alpha_1 \ldots d\alpha_K \\
&=&  SNR^{-d_{\text{out}}(r)}
\end{eqnarray*}
where 
\begin{equation*}
d_{\text{out}}(r) = \inf_{\alpha \in \mathcal{O} } \displaystyle \sum_{k=1}^{K} n_k\alpha_k
\label{eq:optim-asy}
\end{equation*}
The optimisation problem is equivalent to the linear problem in eq.(\ref{eq:LP-2}) where $x_i = n_i \alpha_i$, 
$\hat{\mathbf{x}} = T(\mathbf{x})$, $\hat{\mathbf{n}} = T(\mathbf{n})$ and $\hat{\bar{\boldsymbol{\mu}}} = T(\bar{\boldsymbol{\mu}})$.
\begin{equation}
\begin{cases}
\text{Minimize: } \;\hat x_1 + \ldots + \hat x_K \\
\text{Such that:}\;\;0 \leq \hat x_i \leq \hat n_i \quad \forall i\\ \
\hat{\bar\mu}_1 \hat x_1 + \ldots \hat{\bar\mu}_K \hat x_K \geq 1 - \frac{r}{K} \\ 
\end{cases}
\label{eq:LP-2}
\end{equation}
The optimal solution for this problem is given below 
\\*
If $\underline{r = 0}$, the optimal solution is
$$
\hat{x_i^*} = \hat{n_i} \quad \forall i
$$
If $\underline{r \neq 0}$, then the optimal solution is 
\begin{equation}
\begin{array}{l}
\hat{x_1}^* = \min \Big[ \frac{1}{\hat{\bar{\mu_1}}}\Big(1 - \frac{r}{K}\Big)^+ , \hat{n_1} \Big] \\
\hat{x_i}^* = \min \Big[ \frac{1}{\hat{\bar{\mu_i}}}\Big(1 - \frac{r}{K} - \displaystyle \sum_{j=1}^{i-1}\hat{\bar{\mu_j}}\hat{n_j} \Big)^+, \hat{n_i} \Big] \quad \forall i \geq 2\\
\end{array}
\end{equation}
Then, the DMT is given by
\begin{equation}
d^*_{\text{out}}(r,\mu) = \displaystyle \sum_{i=1}^{K} \hat{x_i^*} \\
\label{eq:DMT-4}
\end{equation}
which is equivalent to the expression of DMT given in (\ref{eq:DMT-asym}). By replacing $r_i$ by its value in (\ref{eq:DMT-asym}), we get the final DMT expression in (\ref{eq:DMT_diff}). 

\end{proof}

\begin{example} We consider 2 weighted parallel ($2\times1$, $1 \times 1$) MISO channels. The DMT with respect to different weight distribution is illustrated in fig. \ref{fig:DMT_asym}. 
If $\mu_1^* = 2/3$, $\mu_2^* = 1/3$ then $d(r,\mu^*) = 3(1 - \frac{r}{2})$ is the upper bound of the DMT. 
\begin{figure}[ht]
\centering
 \includegraphics[width = 0.95\linewidth]{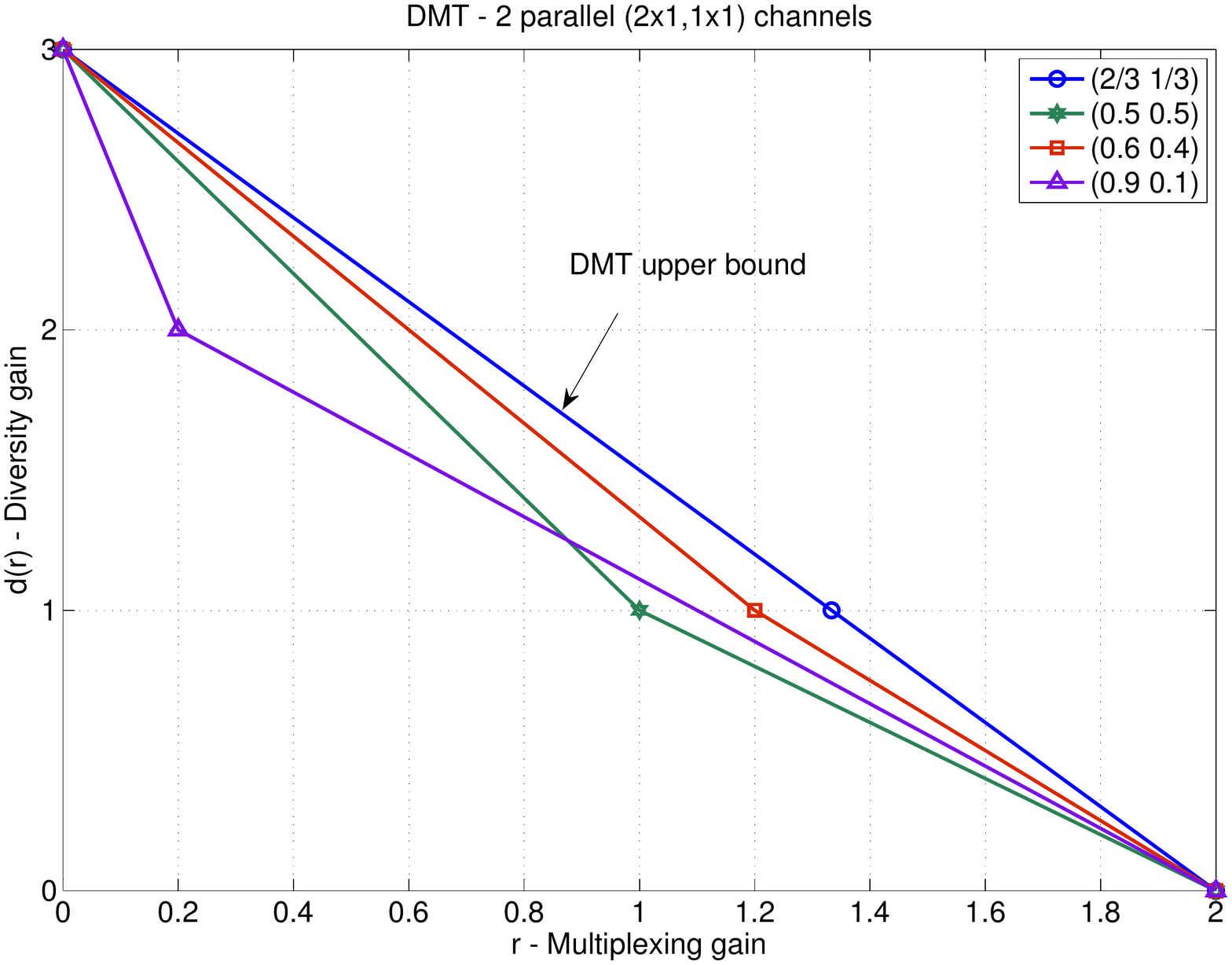}
\caption{DMT of 2 different parallel channels : $2\times 1$ and $1\times 1$ with different weight distributions}
 \label{fig:DMT_asym}
\end{figure}

%\textcolor{blue}{Lina has no idea about what should be in comments, Jean-Claude: commentaire ... }
\end{example}

\section{DMT of the broadcast channel with ZF and DPC precoder}
\label{sec:broadcast}
In this section, we study the DMT of the broadcast channel using ZF and DPC precoder \emph{under fixed weighted sum rate capacity constraint}. 
As shown in section \ref{sec:weighted-sum}, the weighted sum rate capacity of ZF and DPC is asymptotically equivalent to the weighted sum capacity over parallel single-channels. The multiuser MIMO downlink with full CSIT is therefore decomposed into parallel independent single-user channels. We assume that each user has full knowledge of its equivalent channel. Since no cooperation is allowed, every user decodes its own information disregarding the possible dependencies introduced by the precoding schemes.  

\subsection{DMT with ZF and DPC} 
\begin{theorem}
Under fixed weighted sum rate capacity, the DMT of a MIMO broadcast channel with $M$ transmit antennas and K single antenna receivers ($M \geq K$), and using ZF precoder, is a piecewise-linear function connecting the points $(r(i),d(i))$, $i = 0, \ldots, K$, where 
\begin{equation*}
\begin{array}{lll}
r(i) = K\Big(1 - \displaystyle \sum_{j=1}^{K - i} \mu_j\Big) & 0 \leq i \leq  K - 1; & r(k) = K \\
d(i) = (M-K+1)(K-i) & 0 \leq i \leq  K  &\\
\end{array}
\end{equation*} 
\label{theo:DMT_ZF}
\end{theorem}
 
\begin{theorem}
Under fixed weighted sum rate capacity, the DMT of a MIMO broadcast channel with $M$ transmit antennas and K single antenna receivers ($M \geq K$), and using DPC precoder, is a piecewise-linear function connecting the points $(r(i),d(i))$, $i = 0, \ldots, K$, where 
\begin{equation*}
\begin{array}{lll}
r(i) = K\Big(1 - \displaystyle \sum_{j=1}^{K - i} \hat{\mu_j}\Big) & 0 \leq i \leq  K - 1; & r(K) = K \\
d(i) =  \displaystyle \sum_{j=1}^{K - i}\hat{n_j} & 0 \leq i \leq  K-1; &  d(K) = 0 \\
\end{array}
\label{eq:DMT_dpc}
\end{equation*} 
with $\hat{\mathbf{n}} = T(\mathbf{n})$, $\hat{\boldsymbol{\mu}} = T(\boldsymbol{\mu})$ and $n_j = M-j+1$, $\forall j$, when $\mu_1 > \ldots > \mu_K$. Note that, this DMT corresponds to the DPC when the dual uplink order is done in function of increasing weights.
%, \textit{i.e} user $K$ does not get the benefit of any interference cancellation while user 1's signal benefits from all interference cancellation and is thus detected in the presence of the noise only.
\label{theo:DMT_DPC}
\end{theorem}

\begin{proof} Theorems \ref{theo:DMT_ZF} and \ref{theo:DMT_DPC} are immediate consequences of theorems \ref{theo:DMT_ident} and \ref{theo:DMT_dif} respectively (refer to the asymptotical analysis of weighted sum rate capacity in \ref{sec:weighted-sum} for more details).

\end{proof}

\subsection{Numerical example}
\begin{figure}[!ht]
	\centering
		\includegraphics[width= \linewidth]{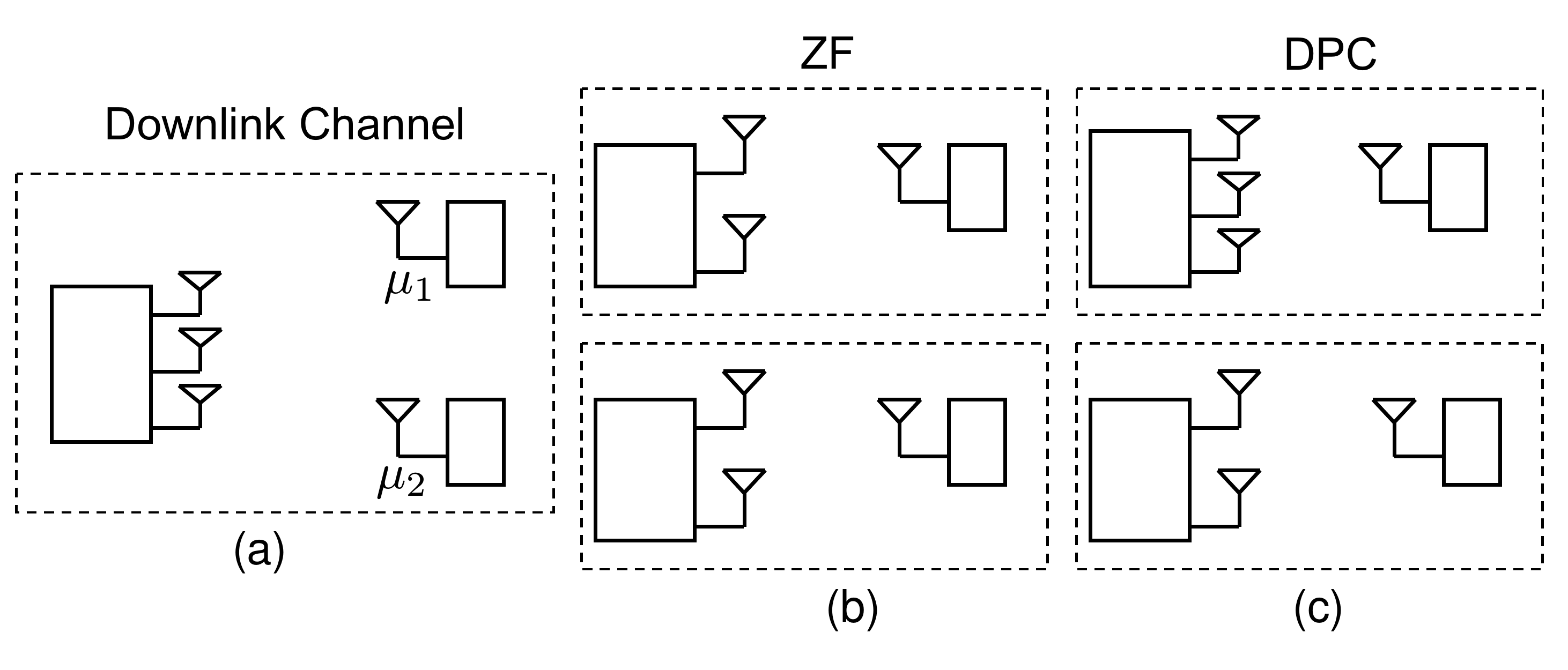}
	\caption{The broadcast channel (a) with $M=3$, $K=2$ and $\mu_1 \geq \mu_2$, can be interpreted in terms of the sum rate of two $2\times 1$ parallel channels when ZF is employed (b), and of $3\times1$ and $2\times1$ parallel channels when DPC is employed (c).}
	\label{fig:eq-dl}
\end{figure}

We consider a broadcast channel with 3 transmit antennas and 2 single antenna users. The parallel equivalent schemes when ZF and DPC precoders are used are illustrated in fig.~\ref{fig:eq-dl}.   

\begin{figure}[!ht]
	\centering
		\includegraphics[width= 0.95\linewidth]{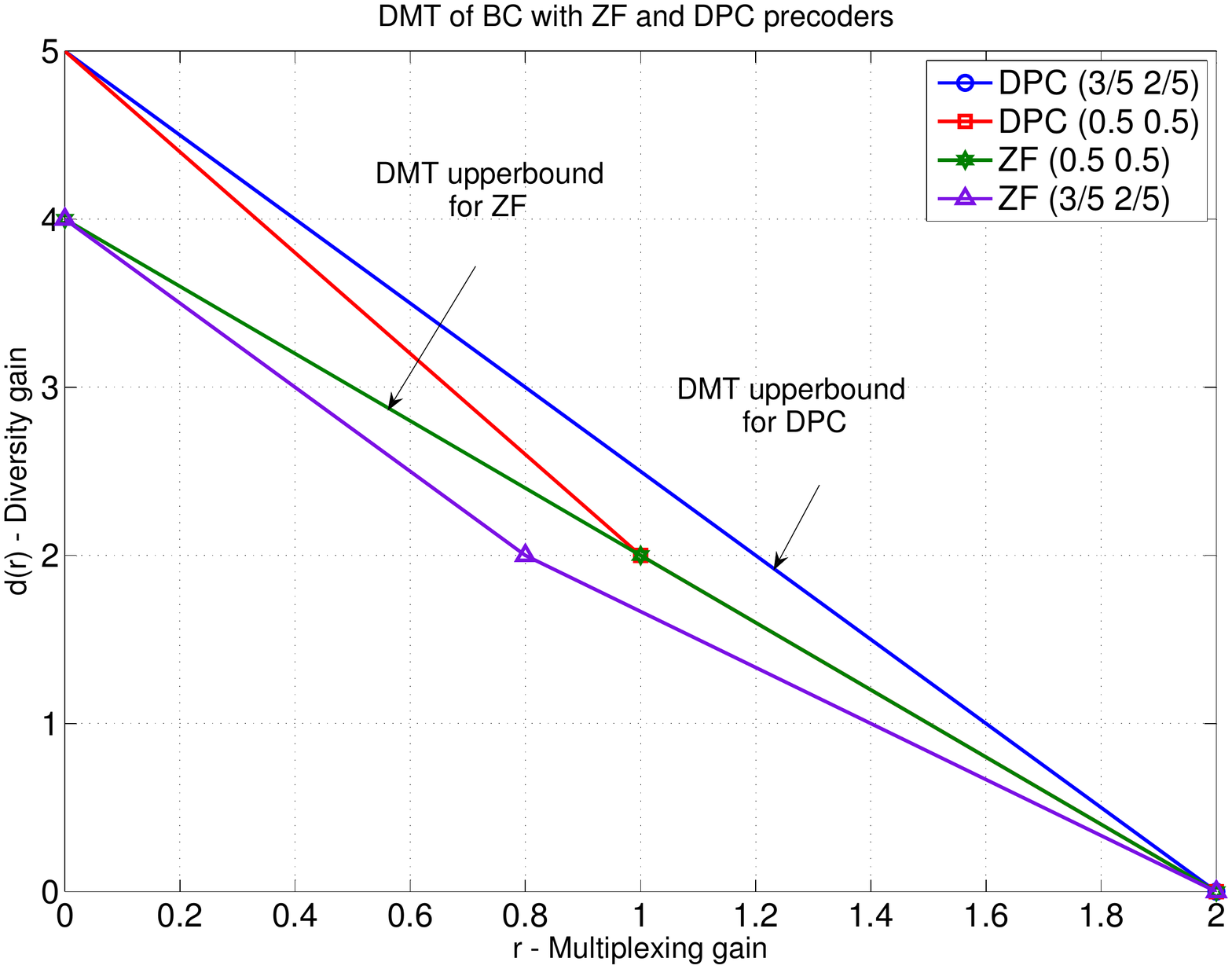}
	\caption{DMT of BC with $M=3$ and $K=2$ for ZF and DPC precoders }
	\label{fig:DMT-BC}
\end{figure}

Fig. \ref{fig:DMT-BC} shows the DMT of ZF versus DPC. We notice that the maximum multiplexing gain ($r = 2$) is the same for both strategies. But the throughput degradation that results from using linear precoding rather than optimal DPC strategies impacts the maximal diversity order ($d_{\text{DPC}} = 5$, $d_{\text{ZF}} = 4$). Moreover, the DMT upperbound of DPC ($\mu_1 = 3/5$, $\mu_2 = 2/5$) is not obtained for the same weight distribution as in the ZF case ($\mu_1 = 0.5$, $\mu_2 = 0.5$).

\section{Conclusion and future work}
\label{sec:conclusion}
In this paper, we derived the DMT of weighted parallel channels when identical and different parallel channels are considered. We show that the DMT depends on the weight distribution. However, the maximal diversity order is independent from the weight distributions. We also found the weight distribution that maximizes the DMT for both cases.
One application is the broadcast channel with Zero Forcing and Dirty paper coding precoding. The DMT of the broadcast channel using these precoders under fixed weighted sum rate capacity constraint has been derived.
 
This work will be extended to the MIMO broadcast case where the receivers have more than one antenna and to the case of imperfect CSIT. These results can be used for a cross-layer study (MAC/PHY). The queuing analysis (MAC layer) optimizes the weight distribution in order to guarantee the stability, whereas the DMT analysis (PHY layer) optimizes the same weight distribution in order to guarantee the reliability of the links. A tradeoff between the two analysis should probably be found in a forthcoming work. 

\section{Acknowledgments}
This work was founded by the Association Nationale de Recherche Technique (ANRT) and by the fond Social Européen (FSE).
 
The authors would like to thank Aitor Del Coso Sanchez from the Centre Tecnologic de Telecomunicacions de Catalunya (CTTC) for his valuable comments on the paper.

\end{document}